\documentclass[%
reprint,
superscriptaddress,
nofootinbib,
amsmath,
amssymb,
aps,
pra,
floatfix,
]{revtex4-2}
\usepackage[table]{xcolor}
\usepackage{graphicx}
\usepackage{dcolumn}
\usepackage{bm}
\usepackage{orcidlink}
\usepackage{booktabs}
\usepackage{optidef}
\usepackage{dsfont}
\usepackage{placeins}
\usepackage[titletoc,toc,page,title,header]{appendix}
\usepackage{ragged2e} 
\usepackage{braket} 
\usepackage{adjustbox}
\usepackage{tabularx}
\usepackage{multirow}
\usepackage[acronym,indexonlyfirst,nomain,nohypertypes={acronym,notation}]{glossaries}
\usepackage{fancyvrb}
\usepackage{mathtools}
\usepackage{relsize}
\usepackage{csquotes}
\usepackage{bbm}
\usepackage{tikz}
\usetikzlibrary{decorations.pathreplacing, calligraphy, positioning, shapes, arrows}
\usepackage[framemethod=TikZ]{mdframed}
\usepackage{siunitx}

\mathtoolsset{showonlyrefs=true}

\hypersetup{
    colorlinks=true,
    linkcolor=blue,
    filecolor=magenta,      
    urlcolor=blue,
    citecolor=blue
}

\definecolor{darkgreen}{HTML}{2C3333}
\definecolor{tabblue}{HTML}{1f77b4}
\definecolor{taborange}{HTML}{ff7f0e}
\definecolor{tabgreen}{HTML}{2ca02c}

\newcommand{\acr}[1]{\hat {\textmd a}_{#1}^{\dagger}}
\newcommand{\aan}[1]{\hat {\textmd a}_{#1}}

\renewcommand{\vec}[1]{\boldsymbol{#1}}

\newcommand{\citer}[1]{Ref.~[\citenum{#1}]}

\newacronym{vrte}{VRTE}{variational real-time evolution}
\newacronym{vite}{VITE}{variational imaginary-time evolution}
\newacronym{nisq}{NISQ}{noisy intermediate-scale quantum}
\newacronym{bo}{BO}{Born–Oppenheimer}
\newacronym{neo}{NEO}{nuclear-electronic orbital}
\newacronym{neohf}{NEO-HF}{nuclear-electronic orbital Hartree–Fock}
\newacronym{hf}{HF}{Hartree–Fock}
\newacronym{uhf}{UHF}{unrestricted Hartree–Fock}
\newacronym{sto6}{STO-6G}{single-$\zeta$ (minimal) basis set contracted from 6 Gaussian primitives}
\newacronym{631}{6-31G}{split-valence double-$\zeta$ Gaussian basis set in 6-31 contraction scheme}
\newacronym{ccd}{cc-pVDZ}{correlation-consistent polarised valence double-$\zeta$}
\newacronym{cc5}{cc-pV5Z}{correlation-consistent polarised valence five-$\zeta$}
\newacronym{pb4}{PB4-F2}{PB-type 4s3p2d2f}
\newacronym{dzsnb}{DZSNB}{split-valence double-$\zeta$ nuclear basis set composed of 2 uncontracted Cartesian $S$ functions}
\newacronym{vqe}{VQE}{variational quantum eigensolver}
\newacronym{avqe}{AdaptVQE}{adaptive variational quantum eigensolver}
\newacronym{neocasci}{NEO-CASCI}{nuclear-electronic orbital complete active space configuration interaction}
\newacronym{casci}{CASCI}{complete active space configuration interaction}
\newacronym{neofci}{NEO-FCI}{nuclear-electronic orbital full configuration interaction}
\newacronym{ci}{CI}{configuration interaction}
\newacronym{fci}{FCI}{full configuration interaction}
\newacronym{ucc}{UCC}{unitary coupled cluster}
\newacronym{neoucc}{NEO-UCC}{nuclear-electronic orbital unitary coupled cluster}
\newacronym{c3h5o2}{C$_3$H$_5$O$_2$}{protonated enol malonaldehyde form}
\newacronym{qpe}{QPE}{quantum phase estimation}
\newacronym{mp2}{MP2}{the second-order Møller–Plesset}
\newacronym{ts}{TS}{transition state}
\newacronym{on}{ON}{occupation number}
\newacronym{cpet}{CPET}{concerted proton-electron transfer}
\newacronym{pt}{PCET}{proton-coupled electron transfer}
\newacronym{isl}{ISL}{incremental structural learning}
\newacronym{aqc}{AQC}{approximate quantum compiling}
\newacronym{adapt-aqc}{ADAPT-AQC}{adaptive approximate quantum compiling}
\newacronym{nfa}{NFA-VQE}{NEO frozen natural orbitals ADAPT-VQE}
\newacronym{zne}{ZNE}{zero noise extrapolation}
\newacronym{fno}{FNO}{frozen natural orbitals}
\newacronym{as}{AS}{active space}
\newacronym{ft}{FTQC}{fault-tolerant quantum computing}
\newacronym{ht}{HAT}{hydrogen atom transfer}

\usepackage{hyperref}

\begin{document}

\title{
Approximate quantum circuit compilation for proton-transfer kinetics on quantum processors
}

\author{Arseny~Kovyrshin\,\orcidlink{0009-0009-1136-4428}} 
\email{arseny.kovyrshin@astrazeneca.com}
\affiliation{Data Science and Modelling, Pharmaceutical Sciences, R\&D, AstraZeneca Gothenburg, Pepparedsleden 1, Molndal SE-431 83, Sweden}
\affiliation{Department of Chemistry and Chemical Engineering, Chalmers University of Technology, Gothenburg, Sweden}

\author{Dilhan~Manawadu\,\orcidlink{0000-0002-3575-8060}}
\affiliation{The Hartree Centre, STFC, Sci-Tech Daresbury, Warrington, WA4 4AD, United Kingdom}

\author{Edoardo~Altamura\,\orcidlink{0000-0001-6973-1897}}
\affiliation{The Hartree Centre, STFC, Sci-Tech Daresbury, Warrington, WA4 4AD, United Kingdom}
\affiliation{Yusuf Hamied Department of Chemistry, University of Cambridge, Lensfield Road, Cambridge CB2 1EW, United Kingdom}

\author{George~Pennington\,\orcidlink{0009-0008-9257-5878}}
\affiliation{The Hartree Centre, STFC, Sci-Tech Daresbury, Warrington, WA4 4AD, United Kingdom}

\author{Benjamin~Jaderberg\,\orcidlink{0000-0001-9297-0175}}
\affiliation{IBM Quantum, IBM Research Europe, Hursley, Winchester, SO21 2JN, United Kingdom}

\author{Sebastian~Brandhofer\,\orcidlink{0000-0002-6010-5643}}
\affiliation{IBM Quantum, IBM Germany Research \& Development, B\"oblingen, Germany}

\author{Anton~Nyk{\"a}nen\,\orcidlink{0009-0001-5849-9508}}
\affiliation{Algorithmiq Ltd., Kanavakatu 3C, FI-00160 Helsinki, Finland}

\author{Aaron~Miller\,\orcidlink{0000-0001-7108-1811}}
\affiliation{Algorithmiq Ltd., Kanavakatu 3C, FI-00160 Helsinki, Finland}
\affiliation{School of Physics, Trinity College Dublin, College Green, Dublin 2, Ireland}

\author{Walter~Talarico\,\orcidlink{0000-0001-6617-7798}} 
\affiliation{Algorithmiq Ltd., Kanavakatu 3C, FI-00160 Helsinki, Finland}
\affiliation{QTF Centre of Excellence, Department of Physics, University of Helsinki, P.O. Box 43, FI-00014 Helsinki, Finland}

\author{Stefan~Knecht\,\orcidlink{0000-0001-9818-2372}}
\affiliation{Algorithmiq Ltd., Kanavakatu 3C, FI-00160 Helsinki, Finland}

\author{Fabijan~Pavo\v{s}evi\'c\,\orcidlink{0000-0002-3693-7546}}
\affiliation{Algorithmiq Ltd., Kanavakatu 3C, FI-00160 Helsinki, Finland}

\author{Alberto~Baiardi\,\orcidlink{0000-0001-9112-8664}}
\affiliation{IBM Quantum, IBM Research Z{\"u}rich, 8803 R{\"u}schlikon, Switzerland}

\author{Francesco~Tacchino\,\orcidlink{0000-0003-2008-5956}}
\affiliation{IBM Quantum, IBM Research Z{\"u}rich, 8803 R{\"u}schlikon, Switzerland}

\author{Ivano~Tavernelli\,\orcidlink{0000-0001-5690-1981}}
\affiliation{IBM Quantum, IBM Research Z{\"u}rich, 8803 R{\"u}schlikon, Switzerland}

\author{Stefano~Mensa\,\orcidlink{0000-0002-0938-144X}}
\affiliation{The Hartree Centre, STFC, Sci-Tech Daresbury, Warrington, WA4 4AD, United Kingdom}

\author{Jason~Crain\,\orcidlink{0000-0001-8672-9158}}
\affiliation{IBM Research Europe, The Hartree Centre, STFC, Sci-Tech Daresbury, Warrington, WA4 4AD, United Kingdom}
\affiliation{Clarendon Laboratory, University of Oxford, Oxford OX1 3PU, UK}

\author{Lars~Tornberg}
\affiliation{Data Science and Modelling, Pharmaceutical Sciences, R\&D, AstraZeneca Gothenburg, Pepparedsleden 1, Molndal SE-431 83, Sweden}

\author{Anders~Broo\,\orcidlink{0000-0001-6121-7818}}
\affiliation{Data Science and Modelling, Pharmaceutical Sciences, R\&D, AstraZeneca Gothenburg, Pepparedsleden 1, Molndal SE-431 83, Sweden}

\date{\today}

\begin{abstract}
Proton transfer reactions are fundamental to many chemical and biological systems, where quantum effects such as tunneling, delocalization, and zero-point motion play key kinetic control roles. However, classical methods capable of accurately capturing these phenomena scale prohibitively with system size. Here, we develop and demonstrate quantum computing algorithms based on the Nuclear-Electronic Orbital framework, treating the transferring proton quantum mechanically. We assess the potential of current quantum devices for simulating proton transfer kinetics with high accuracy. We first construct a deep initial ans\"atze within a truncated orbital space by employing the frozen natural orbital approximation. Then, to balance circuit depth against state fidelity, we implement an adaptive form of approximate quantum compiling. Using resulting circuits at varying compression levels transpiled for the \texttt{ibm\_fez} device, we compute barrier heights and delocalised proton densities along the proton transfer pathway using a realistic hardware noise model. We find that, although current quantum hardware introduces significant noise relative to the demanding energy tolerances involved, our approach allows substantial circuit simplification while maintaining energy barrier estimates within 13\% of the reference value. Despite present hardware limitations,  these results offer a practical means of approximating key circuit segments in near-term devices and early fault-tolerant quantum computing systems. 
\end{abstract}

\maketitle

\section{Introduction}
\label{sec:intro}

Proton transfer plays a crucial role in many physical and biochemical processes. For example, several enzymes, particularly those involved in metabolic pathways, use proton transfer as part of the catalytic cycle~\cite{liu2021dual}.
Kinases~\cite{liu2021reactivities}, oxidoreductases~\cite{silva2022alternative}, and dehydrogenases~\cite{stripp2022second} rely on the transfer of protons between the residues of the active site to catalyze reactions.  Coupled proton and electron transfers are known to be important for drugs targeting mitochondrial function or redox-based mechanisms~\cite{murgida2001proton}, and their rates can influence the overall reactivity and efficiency of drug candidates targeting redox enzymes. Similarly, proton transfer can be the rate-determining step in enantioselective reactions involving chiral substrates~\cite{cao2021catalytic}. 
 
Typically, the reaction rates $k(T)$ of these proton transfer processes at a temperature $T$ are related to the barrier height $E_b$ by an Arrhenius law $k(T) \propto e^{-E_b/RT}$, which assumes that the mobile nuclei behave classically~\cite{atkins2010physical}.  Given such exponential sensitivities, accurate prediction of the proton transfer energy barriers involved is critical for rate control and in the design of reactions that favor one pathway over another. However, a significant complexity arises because proton transfer often involves light nuclei for which the classical approximation fails, and non-Arrhenius behavior arises from quantum mechanical phenomena such as zero-point energy, nuclear delocalization and tunneling.  Consequently, reactions proceed at rates inconsistent with classical transition state theory. In these circumstances, treating the labile protons quantum mechanically is necessary to recover physically realistic models.

Several theoretical approaches have been developed over the past decades that can accurately describe proton transfer~\cite{habershon2013ring,beck2000multiconfiguration}; however, they are computationally demanding and therefore impractical for large-scale applications. The \gls{neo} approach~\cite{webb2002multiconfigurational,pavosevic2020chemrev}, on the other hand, has emerged as a computationally efficient alternative for capturing the proton transfer process. In the NEO method, both electrons and selected transferring protons are treated quantum mechanically with a molecular orbital theory approach. Therefore, the NEO framework offers an ideal computational platform for the simulation of proton transfer processes~\cite{zhao2020real,zhao2021excited,tao2021direct,dickinson2023generalized,dickinson2024nonadiabatic}. These simulations typically rely on multicomponent density functional theory (DFT) mean-field method~\cite{pak2007density,yang2017development,brorsen2017multicomponent} either in the context of real-time NEO time-dependent density functional theory (RT-NEO-TDDFT)~\cite{zhao2020real} and RT-NEO-TDDFT Ehrenfest dynamics~\cite{zhao2021excited} or multi-state NEO density functional theory (NEO-MSDFT)~\cite{yu2020nuclear}. Due to the well-known limitations of currently available electron-electron~\cite{cohen2012challenges} and electron-proton functionals~\cite{yang2017development,brorsen2017multicomponent}, these methods do not provide adequate accuracy to meet practical, experimental demands (below 1~kcal/mol), providing the impetus for the development of NEO methods of much higher accuracy.  

On classical computers, the resources required for exact solutions of NEO-based models grow exponentially with system size~\cite{pavosevic2020chemrev}. By contrast, advances in quantum computing technology and algorithm developments have opened a path for solving this problem nearly exactly on quantum devices in a more efficient way~\cite{aspuru2005simulated,veis2016quantum,cao2019quantum,bauer2020quantum,kovy2023jpcl}. As \gls{ft} remains a prospect, algorithms tailored to address the challenges posed by noisy intermediate-scale quantum (NISQ)~\cite{preskill2018quantum} hardware are currently being developed. 

In this work, we extend the recently developed NEO framework for modelling proton transfer reactions~\cite{kovy2023jpcl}. First, we enhance the physical realism of the model by employing larger basis sets, the \gls{fno} approach, and by accounting for scaffold relaxation effects. Second, we use the Adaptive Derivative-Assembled Pseudo-Trotter ansatz Variational Quantum Eigensolver (ADAPT-VQE) algorithm for efficient wavefunction parametrization~\cite{nyke2023}, followed by circuit compression using \gls{aqc}~\cite{khatri2019quantum, sharma2020noise, jaderberg2022solving}, enabling applications on \gls{nisq} hardware. Third, the resulting circuits, at varying levels of compression, are then used to simulate adiabatic proton transfer in the malonaldehyde molecule~\cite{kovy2023jpcl}, with a focus on computing the transfer barrier in the presence of quantum nuclei.

\section{Theory}
\label{sec:theory}
We investigate the quantum dynamics of proton transfer in malonaldehyde using a time-dependent quantum simulation framework. Rather than a static description, this approach captures the evolution of both electronic and nuclear degrees of freedom, guided along a well-defined, chemically relevant proton transfer pathway. The core of our approach is to solve the time-dependent Schr\"odinger equation (TDSE) for the coupled proton-electron system, leveraging the NEO framework. In Section \ref{sec:theory:proton-dynamics} we describe the TDSE without going into details regarding the actual molecular setup, which will be covered in Sec. \ref{sec:theory:malonaldehyde}.

\subsection{Coupled electron-proton dynamics}
\label{sec:theory:proton-dynamics}
 The dynamics of the system is governed by the TDSE
\begin{equation}
    \label{eq:tdS}
    i \hbar~\frac{\partial}{\partial t} \big|\Psi(t)\big\rangle = \left[ \hat T_{\text{p}} + \hat V(t) \right]~\big|\Psi(t)\big\rangle \, ,
\end{equation}
where $\hat{T}_{\text{p}}$ is the kinetic energy operator for the proton, and $\hat{V}(t)$ is a time-dependent potential incorporating both the proton's interaction with its environment and the coupling between protonic and electronic components. The explicit time dependence in $\hat{V}(t)$ arises from our protocol for steering the system Hamiltonian along the reaction pathway, as described in Sec.~\ref{sec:theory:malonaldehyde}. 
The total time-dependent potential is given by
\begin{equation}
    \label{eq:Vp}
    \hat V(t) = \hat V_{\text{p}}(t) + \hat V_{\text{ep}}(t) + \hat H_{\text{e}}(t) \, 
\end{equation}
where $\hat V_{\text{p}}(t)$ describes the time-dependent interaction of the proton with the classical nuclear scaffold,
$\hat V_{\text{ep}}(t)$ is the proton-electron interaction, and 
$\hat H_{\text{e}}(t)$ defines the electronic Hamiltonian contribution.
Specifically, the proton-classical-nuclear interaction is represented in second quantization as
\begin{equation}
    \hat V_{\text{p}}(t) = \sum_{PQ} v_{PQ}(t) \acr{P} \aan{Q}\, ,
\end{equation}
The coupling between protonic and electronic degrees of freedom is written as
\begin{equation}
    \hat V_{\text{ep}}(t) = \sum_{PQpq} g_{PQpq}(t)~\acr{P} \acr{p} \aan{q}\aan{Q}, 
\end{equation}
and the purely electronic part is given by
\begin{equation}
    \hat H_{\text{e}}(t) = \sum_{pq} h_{pq}(t)~ \acr{p} \aan{q} 
    +\frac{1}{2}\sum_{pqrs} h_{pqrs}(t)~ \acr{p} \acr{r} \aan{s} \aan{q} \, .
\end{equation}
In this work, the lower case $p, q, r, s$ indices denote
general electronic spin orbitals, whereas the
upper case indices denote protonic spin orbitals. The operators \(\acr{p}\) and \(\aan{p}\) are creation and annihilation operators which satisfy the normal fermionic anticommutation relations. 
The explicit time dependence of integrals $v_{PQ}(t)$, $g_{PQpq}(t)$, $h_{pq}(t)$, and $h_{pqrs}(t)$ arises from the fact that the underlying electronic molecular orbitals are defined with reference to a representative proton transfer pathway (see Sect.~\ref{sec:theory:malonaldehyde} for more details). This prescribed evolution of the electronic orbitals steers the quantum dynamics of both the electronic and protonic degrees of freedom. 

For the time-dependent wave function with protonic (nuclear) and electronic components $\ket{\Phi_{\nu}^{\text{p}}}$ and $\ket{\Phi_{\mu}^{\text{e}}}$ respectively, we write
\begin{align}
    \label{eq:neofci}
    \ket{\Psi [\vec{C}(t)]} = \sum_{\mu \nu}C_{\mu \nu}(t) \ket{\Phi_{\mu}^{\text{e}}} \ket{\Phi_{\nu}^{\text{p}}} \ , 
\end{align}
which inserted into Eq.~\eqref{eq:tdS} leads to an equation of motion for the \acrfull{ci} coefficients $\vec{C}(t)$
\begin{equation}\label{eq:motion}
    \vec{H}(t) \vec{C}(t) = i\frac{\partial}{\partial t}\vec{C}(t)\, ,
\end{equation}
where the matrix elements of $\vec{H}(t)$ are defined by
\begin{equation}\label{eq:Ht}
    H_{\kappa \lambda, \mu \nu}(t)=\bra{\Phi^{\text{e}}_{\kappa}}\bra{\Phi^{\text{p}}_{\lambda}} \hat T_{\text{p}} + \hat V(t)\ket{\Phi^{\text{e}}_{\mu}} \ket{\Phi^{\text{p}}_{\nu}} \, .
\end{equation}

The rate of change of $\hat V(t)$ determines whether the system is in the adiabatic or non-adiabatic regime. Previously, Kovyrshin \textit{et al.} (2023)~\cite{kovy2023jpcl} explored dynamics through Suzuki decomposition in both cases. While highly accurate, this approach requires quantum circuits of substantial depth for time evolution---a limitation for near-term quantum hardware.

To address these computational challenges, we adapt our protocol for compatibility with variational quantum algorithms, specifically the variational quantum eigensolver (\gls{vqe})~\cite{peru2014}. 
Rather than simulating explicit real-time evolution according to Eq.~\eqref{eq:motion}, we approximate the time-dependent energy and wavefunction variationally at each point along the reaction pathway,  
that is, at each time, the energy and ground state of the instantaneous proton-electron Hamiltonian is obtained by minimizing
\begin{equation}
    \label{eq:AdTr}
    E(t) = \min_{\vec{C}(t)} \frac{\bra{\Psi [\vec{C}(t)]} \hat{T}_{\text{p}} + \hat{V}(t) 
    \ket{\Psi[\vec{C}(t)]}}{\bra{\Psi [\vec{C}(t)]}\ket{\Psi [\vec{C}(t)]}}.
\end{equation}
Although limited to the adiabatic regime, this strategy allows for quantum simulation of proton transfer using compact variational ansätze of nearly constant size, making the problem tractable for near-term devices while still capturing such quantum effects as zero-point energy and proton delocalization.

\subsection{Malonaldehyde setup}
\label{sec:theory:malonaldehyde}
The process of \gls{pt} in malonaldehyde, a prototypical model for enantioselective isomerization, has been previously studied by Kovyrshin \textit{et al.}~\cite{kovy2023jpcl} in adiabatic and nonadiabatic limits.  
That study served as a proof-of-concept demonstration, which was limited to near-minimal electronic and nuclear basis sets. In this work, we focus on enhancing the physical realism of the model to obtain quantitatively accurate results comparable to experimental measurements. This is done by employing a larger basis set for electronic and protonic orbitals (see Appendix~\ref{sec:basis}), allowing scaffold relaxation, and leveraging the \gls{fno} technique~\cite{nyke2023}.  Next, we describe the procedures for generating the set of molecular orbitals needed to represent the system and for modelling the proton transfer through the adiabatic steering of the system Hamiltonian. 

Similarly to~\cite{kovy2023jpcl}, we describe the dynamics of proton transfer by first generating the structure of the molecule at three stationary points along a predefined reaction path. These describe the initial (\textit{Left}), transition (\textit{Middle}) and final (\textit{Right}) states respectively.  These structures and their corresponding molecular orbitals, which are needed to model the system, are obtained in three steps. 

First, the three possible positions for the H atom and the corresponding scaffolds are modelled within the Born-Oppenheimer approximation, treating all nuclei as classical point particles. This is done using a \gls{mp2} theory, which yields the fully optimised structure under the $C_s$ point-group symmetry (see Appendix~\ref{sec:basis} for basis set specifications). The corresponding geometries are provided in Table~\ref{tab:xyz_all} in Appendix~\ref{sec:xyz}. Thus, in contrast to~\cite{kovy2023jpcl}, our setup includes scaffold relaxation upon proton transfer. 
The resulting proton positions (see blue points in Fig.~\ref{fig:TOC} and Table~\ref{tab:pos} in Appendix~\ref{sec:xyz}) serve as a good initial guess for the subsequent step where we go beyond the Born–Oppenheimer approximation by treating the transferring proton quantum mechanically.
Second, we perform one \gls{neohf} calculation per scaffold, keeping the protonic orbitals centered at these positions.
This step yields the occupied protonic and electronic molecular orbitals for the system. Third, to obtain more accurate virtual orbital space (\textit{i.e.,} unoccupied orbitals), a higher accuracy method --- \gls{neo}-\gls{mp2} --- is used to generate the corresponding \acrshort{neo}-\acrshort{fno} orbitals (see Sec.~\ref{sec:fno-adat} for more details). 

The construction of the orbitals for the \textit{Right}, \textit{Middle} and \textit{Left} configurations enables us to define a Hamiltonian which represents a changing external potential in which the proton under study evolves. In order to describe this situation, we construct the three Hamiltonian operators in a second quantization based on the sets of protonic and electronic orbitals derived above. The protonic \gls{as} was chosen to contain two orbitals from \textit{Left}, \textit{Middle}, and \textit{Right} setups, and is shared among the three Hamiltonians. Specifically, one occupied \gls{neohf} and one \gls{fno} orbital from each setup result in 6 protonic orbitals. As these orbitals originate from separate calculations, they are not orthogonal, and a L\"owdin orthogonalization procedure~\cite{loew1950} was performed before 
inclusion of these to protonic \gls{as}. While the same protonic \gls{as} was used for all Hamiltonians, the three different electronic \gls{as} 
comprised of 3 \gls{neohf} and 5 \gls{fno} orbitals from \textit{Left}, \textit{Middle}, and \textit{Right} configurations (the 16 lowest
occupied orbitals were considered frozen).  Thus, the protonic and electronic active spaces were doubled in size compared to~\citer{kovy2023jpcl}, and six electrons were considered in the active space instead of four. A schematic representation of the active spaces is shown in the lower part of Fig.~\ref{fig:TOC}.

First, we benchmark the combined effect combined methodological improvements by comparing \gls{casci} energies computed for the \textit{Left} and \textit{Middle} Hamiltonians using our current approach, \gls{fno}-\gls{neo}-\gls{casci}, with the results from~\citer{kovy2023jpcl}, referred to as \gls{neo}-\gls{casci}. In Table~\ref{tab:neo-fno-fci} we list the calculated energy barrier, $\Delta E = E_\mathrm{M}-E_\mathrm{L/R}$, for the two methods along with the experimental and semi-empirical values for comparison. We find that \gls{fno}-\gls{neo}-\gls{casci} doubles the proton transfer barrier compared to~\citer{kovy2023jpcl}, yielding a value of 11.9 \textrm{mHa}. This value falls between the experimentally inferred estimate of approximately 13.5 \textrm{mHa}~\cite{smit1983} and the semi-empirical model-derived value of approximately 11.2 \textrm{mHa}~\cite{fill2005}.  
The former is based on a broad IR absorption near 2960 \textrm{cm}$^{-1}$ tentatively attributed to excitation along the proton transfer coordinate. 
Similarly, we observe an effect on proton–electron entanglement within the system. The entanglement entropy is calculated from the single-proton reduced density matrix, obtained by tracing out the electronic degrees of freedom, for the \textit{Left}, \textit{Middle}, and \textit{Right} ground states. As shown in Table~\ref{tab:neo-fno-fci}, a considerable increase in entanglement is observed for the \textit{Middle} state, while \textit{Left} and \textit{Right} states show only minor improvements. This indicates that the new framework, \gls{fno}-\gls{neo}-\gls{casci}, is better suited to capturing proton-electron entanglement compared to \gls{neo}-\gls{casci}~\cite{kovy2023jpcl}.

To asses the spatial delocalization arising from the quantum treatment of the proton, we analyze the protonic density and deviation from classical reference position. Specifically, we evaluate the protonic densities and the expectation values of the proton position operator from a protonic one-particle density matrix constructed for \textit{Left}, \textit{Middle}, and \textit{Right} FNO-NEO-CASCI states (see Fig.~\ref{fig:TOC}). The corresponding proton densities reveal a strongly delocalised character across all three setups, emphasizing the inadequacy of a purely classical treatment. The expectation values for the proton position (white points in Fig.~\ref{fig:TOC} and Table~\ref{tab:pos}) exhibit substantial displacement from the original centres of protonic basis functions -- i.e., the classical reference positions. These findings highlight the importance of quantum nuclear effects where proton motion plays a central role. 

With the \textit{Left}, \textit{Middle}, and \textit{Right} Hamiltonians prepared, we now define the time-dependent Hamiltonian for the system
\begin{equation}
\label{eq:HLMRT}
    H(t) = \alpha(t) H_{\rm Left} + \beta(t) H_{\rm Middle} + \gamma(t) H_{\rm Right}.
\end{equation}
This is then used in Eq.~\eqref{eq:Ht} to adiabatically change the system from $H_\mathrm{Left}$ to $H_\mathrm{Middle}$ to $H_\mathrm{Right}$, by an appropriate choice of the parameters $\alpha$, $\beta$ and $\gamma$. This thus defines the adiabatic trajectory for proton transfer, as governed by 
Eq.~\eqref{eq:AdTr} and Eq.~\eqref{eq:motion}, where the \textit{Left}, \textit{Middle}, and \textit{Right} energies can be considered as stationary points along 
this trajectory. 
\begin{figure*}
    \centering
    \includegraphics[width=0.8\linewidth]{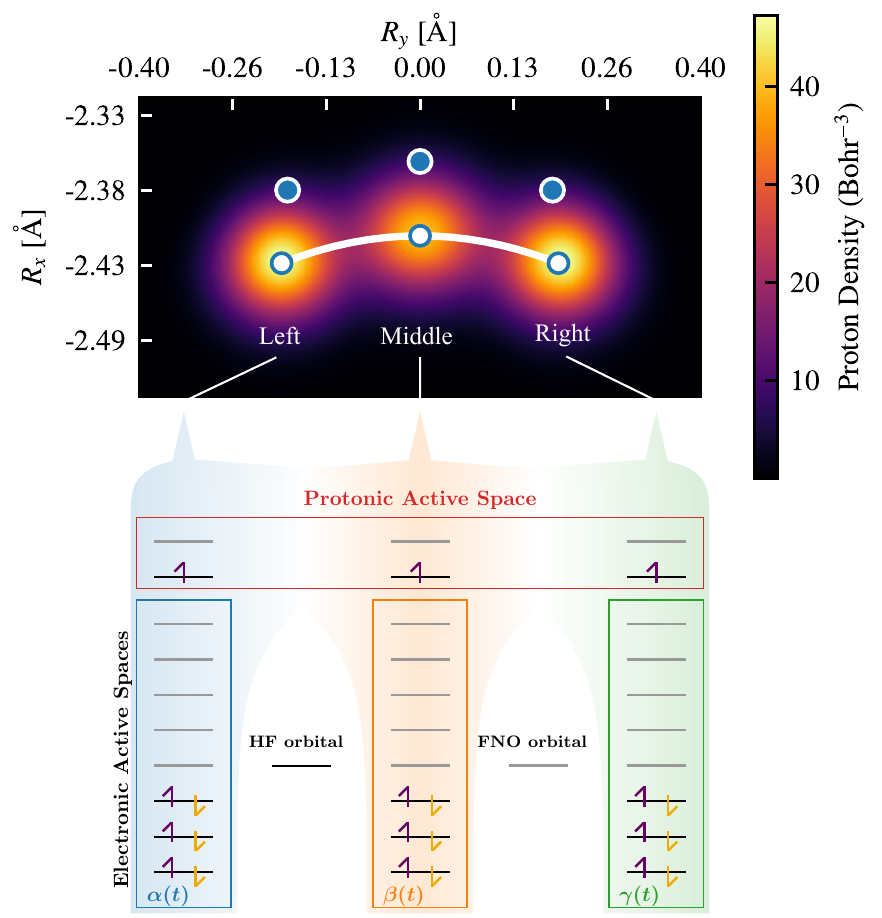}
    \caption{Proton density distribution for \textit{Left}, \textit{Middle}, and \textit{Right} setups computed from the FNO-NEO-CASCI protonic one-particle density matrix. The proton density extends downward, and the classical nuclei lie above the plotted region. White points indicate the expectation values of the proton position operator for \textit{Left}, \textit{Middle}, and \textit{Right} setups, while blue points denote the positions of the protonic orbital centers used in the basis set expansion. The overlaid white trajectory indicates the expectation value path of the proton position. Schematic illustration of the electronic and protonic active spaces used in the \textit{Left}, \textit{Middle}, and \textit{Right} Hamiltonians. Gray lines indicate the FNO orbitals, and black lines represent the HF orbitals. For each setup, the occupation of the lowest energy product state (HF state) is shown.}
    \label{fig:TOC}
\end{figure*}
\begin{table}
    \begin{tabular}{@{}lccc@{}}
    \toprule
    Method & $\Delta E$& \multicolumn{2}{c}{Entanglement entropy} \\
           &  (mHa)    & \textit{Left}/\textit{Right} & \textit{Middle} \\
    \midrule
    NEO-CASCI$^\star$    & 5.1 & 0.00200 & 0.0038 \\
    FNO-NEO-CASCI        & 11.9 & 0.00240 & 0.0066 \\
    Semi-empirical model~\cite{fill2005}     & 11.2 &         &        \\ 
    Experimental~\cite{smit1983}          & 13.5 &         &        \\  
    \bottomrule
    \end{tabular}
    \caption{NEO-CASCI reference energies from~\citet{kovy2023jpcl} ($^\star$), and FNO-NEO-CASCI reference energies with entanglement entropy values.}
    \label{tab:neo-fno-fci}
\end{table}
\subsection{Transition rate constant}
To highlight the importance of accurately estimating the reaction barrier, it is useful to demonstrate its effect on reaction kinetics. The temperature-dependent rate constant for proton transfer, incorporating quantum nuclear effects via the NEO method, can be computed using a Transition State Theory (TST) expression~\cite{atkins2010physical}
\begin{equation}
\label{eq:Eyring}
k(T) \propto  \exp \left( -\frac{\Delta E}{k_BT} \right),
\end{equation}
where $T$ is the temperature and \( k_B \) is Boltzmann’s constant.
This formulation parallels the Eyring equation~\cite{atkins2010physical} but uses an effective quantum barrier, \( \Delta E \), instead of Gibbs free energy activation.
The barrier incorporates proton zero-point energy and proton delocalization effects, making the expression particularly suitable for the light-particle transfer process. Although thermal and entropic contributions are omitted, this is a reasonable approximation given their relatively minor role in the short-timescale proton transfer process. 

\section{Methods}
\label{sec:methods}
The methods presented in this section detail the computational pipeline used to simulate quantum proton transfer dynamics in malonaldehyde. We begin with a schematic overview of our multi-stage workflow, which integrates accurate \textit{ab initio} reference calculations, variational quantum algorithms, circuit compression, and error mitigation techniques. Subsequent subsections provide in-depth descriptions of the ADAPT-VQE method for quantum wavefunction preparation, frozen natural orbital concept, the Adaptive Approximate Quantum Compiling (ADAPT-AQC) approach for circuit depth reduction, and the implementation of Zero-Noise Extrapolation (ZNE) to address the effects of hardware noise in realistic quantum simulations.
\subsection{Pipeline overview}

\label{sec:methods:overview}

\begin{figure*}
\centering

\begin{tikzpicture}[
    node distance=1.6cm, auto,
    block/.style={rectangle, draw, rounded corners, minimum height=2.5em, minimum width=5em, text centered, fill=blue!10},
    decision/.style={diamond, draw, minimum width=3cm, minimum height=1cm, text centered, text width=3cm, fill=yellow!20},
    cloud/.style={ellipse, draw, minimum width=3cm, minimum height=1cm, text centered, text width=3cm, fill=green!20},
    line/.style={draw, line width=0.75mm, -latex', black},
    connector/.style={draw, circle, fill=red!20, text width=2em, text centered, node distance=4cm}
  ]
    \node [block] (casci) {FNO-NEO-CASCI (ground truth)};
    \node [block, below of=casci, style={fill=red!20}] (vqe) {Ground state preparation};
    \node [block, right of=vqe, xshift=3.5cm] (vqedeep) {ADAPT-VQE (deep)};
    \node [block, below of=vqe] (vqeshallow) {ADAPT-VQE (shallow)};

    \node [block, below of=vqeshallow, style={fill=red!20}] (aqc) {Circuit compiling};
    \node [block, below of=aqc] (aqcdeep) {ADAPT-AQC (high)};
    \node [block, right of=aqc, xshift=3.5cm] (aqcshallow) {ADAPT-AQC (low)};
    
    \node [cloud, below of=aqcdeep] (Statevector) {Statevector\\ (early-FT regime)};
    \node [cloud, below of=aqcshallow] (NoiseSim) {Noisy simulation (\texttt{ibm\_fez})};
    \node [block, below of=NoiseSim] (ZNE) {Error mitigation with ZNE};

    \path [line] (casci) -- (vqe);
    \path [line] (vqe) -- (vqeshallow);
    \path [line] (vqe) -- node[anchor=south] {} (vqedeep);
    \path [line] (vqeshallow) -- (aqc);
    \path [line] (aqc) -- (aqcshallow);
    \path [line] (aqc) -- (aqcdeep);
    \path [line] (aqcdeep) -- (Statevector);
    \path [line] (aqcshallow) -- (NoiseSim);
    \path [line] (NoiseSim) -- (ZNE);
\end{tikzpicture}

\caption{Schematic of our multi-stage proton-transfer simulation workflow. We begin with the high-precision FNO-NEO-CASCI reference (blue block) and perform ADAPT-VQE optimization (red block). By running the VQE driver to two distinct convergence thresholds, we generate both \textit{deep} and \textit{shallow} ADAPT-VQE circuits. These are then passed through Adaptive Approximate Quantum Compiling (ADAPT-AQC, red block), yielding \textit{high-} and \textit{low-}fidelity contracted circuits with significantly reduced two-qubit depth compared to ADAPT-VQE shallow. The ADAPT-AQC deep circuit is simulated in the noiseless regime using a state-vector simulator, while the ADAPT-AQC shallow circuit is executed on the noisy \texttt{ibm\_fez} model combined with Zero-Noise Extrapolation (ZNE) error mitigation.}
\label{fig:workflow}
\end{figure*}
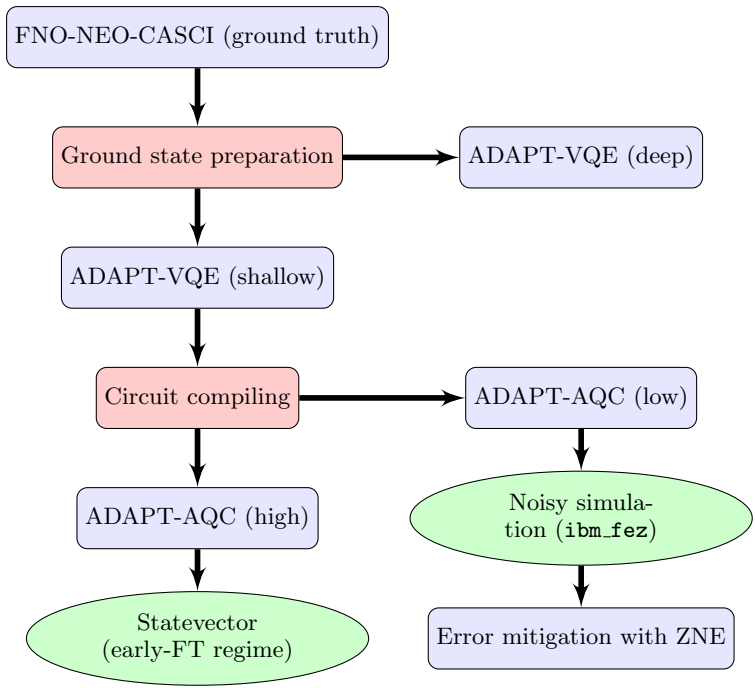

The computational workflow for simulating quantum proton transfer dynamics in malonaldehyde is illustrated in Fig.~\ref{fig:workflow}. Our pipeline integrates high-level \textit{ab initio} electronic structure theory with state-of-the-art quantum algorithms, ensuring both accuracy and practical feasibility for near-term quantum hardware. The steps are as follows.

First, we compute ground-state energies using the FNO-NEO-CASCI approach described in Section~\ref{sec:theory:malonaldehyde}, which we refer to as CASCI in the following. This provides a reliable theoretical reference for benchmarking proton transfer barriers and quantum simulation outcomes.

Next, the ADAPT-VQE (see Sec. \ref{sec:fno-adat} for details) is used to prepare quantum circuits representing the ground-state wavefunction for the time-dependent Hamiltonian at different times. This procedure is carried out at two distinct accuracy thresholds: an energy error of $10^{-3}$ \textrm{Ha} and a less conservative value of $10^{-2}$ \textrm{Ha}, relative to the CASCI reference energy. The selected convergence threshold affects the required circuit depth, balancing computational accuracy and quantum hardware feasibility. The resulting deep circuit, VQE-deep, will be used as a benchmark to represent the fault tolerant regime, whereas the shallow circuits, VQE-shallow, will be further compressed as described below. 
To ensure compatibility with near-term quantum devices, where circuit depth and gate error rates are critical constraints,  we further compress the ADAPT-VQE circuits utilizing \gls{adapt-aqc}~\cite{jaderberg2025variational}. This procedure significantly reduces circuit depth at the cost of a further accuracy trade-off. The method provides control over the level of compression tailored to different hardware capabilities. In our specific case, we produce both high- and low-fidelity compiled circuits, denoted AQC-high and AQC-low.

The compiled circuits are subsequently simulated in two regimes: (1) the high-fidelity circuit is used in an ideal, noiseless statevector simulator (representative of the early fault-tolerant quantum era), and (2) the low-fidelity circuit is used under realistic noise models based on the \texttt{ibm\_fez} Heron processor, which includes comprehensive device imperfections (see Appendix~\ref{sec:noise} for details). For the latter, we apply Zero-Noise Extrapolation (ZNE), an error mitigation protocol, to recover expectation values closer to the noise-free limit.

The following sections provide further details for the ADAPT-VQE, FNO, ADAPT-AQC, and ZNE techniques.

\subsection{ADAPT-VQE with NEO-FNO pool}\label{sec:fno-adat}

At the core of our quantum simulation protocol is the ADAPT-VQE algorithm~\cite{nyke2023}, applied to a pool of the NEO excitation operators spanning the space defined by FNO approach. It is used to approximate the ground state wavefunction $|\Psi[\vec{C}(t)]\rangle$ for each time-dependent Hamiltonian along the adiabatic trajectory.
This method, detailed in ~\citer{nyke2023}, delivers a wave function with a compact circuit and high accuracy; we briefly review it below. As an extension of its electronic counterpart~\cite{grimsley2019adaptive}, it bears similar features: the wavefunction ansatz is built iteratively by adding operators from a pool comprising single and double electronic, protonic, and mixed fermionic excitations: 
\begin{equation}
\tau^\mu\in\{a_{i}^{a}-a^{i}_{a},a_{I}^{A}-a^{I}_{A},a_{ij}^{ab}-a^{ij}_{ab},a_{IJ}^{AB}-a^{IJ}_{AB},a_{iI}^{aA}-a^{iI}_{aA}\}, 
\end{equation}
where $a_{i}^{a}=(a^{i}_{a})^{\dagger}=a^{\dagger}_aa_i$, $a_{I}^{A}=(a^{I}_{A})^{\dagger}=a^{\dagger}_Aa_I$ are single electronic and protonic, respectively, excitation operators, and $a_{ij}^{ab}=(a^{ij}_{ab})^\dagger=a^{\dagger}_aa^{\dagger}_ba_ja_i$, $a_{IJ}^{AB}=(a^{IJ}_{AB})^\dagger=a^{\dagger}_Aa^{\dagger}_Ba_Ja_I$, and $a_{iI}^{aA}=(a^{iI}_{aA})^\dagger=a^{\dagger}_aa^{\dagger}_Aa_ia_I$ are double electron-electron, proton-proton, and electron-proton excitation operators. Here, the lower-case $i,j,k,l,...$, $a,b,c,d,...$ indices stand for occupied and virtual electronic spin orbitals, respectively. The protonic spin orbitals are defined analogously using upper-case indices. At each iteration, the operator that most effectively lowers the system energy is selected and added to the variational circuit together with its optimised parameter. After the $n$-th iteration, the ADAPT-VQE ansatz thus takes the form
\begin{equation}
\label{eq:neo-adapt-ansatz}
|\Psi_{\text{ADAPT-VQE}}^{(n)}\rangle=e^{\theta_n\tau^n}...e^{\theta_2\tau^2}e^{\theta_1\tau^1}|0^{\text{e}}0^{\text{p}}\rangle
\end{equation}
where ($|0^{\text{e}}0^{\text{p}}\rangle=|0^{\text{e}}\rangle\otimes|0^{\text{p}}\rangle$) is the reference \gls{neohf} state composed from the electronic ($|0^{\text{e}}\rangle$) and protonic ($|0^{\text{p}}\rangle$) Slater determinants, and $\theta_n$ are the ans\"atze parameters. The ansatz grows until the convergence defined by a tolerance threshold is achieved. Further reduction in the circuit depth is achieved by selecting the operators from the qubit pool instead of the fermionic pool. The qubit pool of operators is constructed by mapping the fermionic operator pool to the qubit space and splitting the operators into individual strings~\cite{jordan1993algebraic,bravyi2002fermionic,bonsai}.

\subsubsection{Frozen Natural Orbitals Approximation}
To account for the missing dynamical correlation between quantum particles, we utilise the FNO approximation. The FNO approximation provides a means for systematic truncation of the unoccupied orbitals without sacrificing accuracy~\cite{sosa1989selection}. As previously introduced in Ref.~\citenum{nyke2023}, the electronic and protonic FNOs are defined as eigenvectors of a one-particle electronic density matrix
\begin{equation}
    \label{eqn:electronic-1rdm}
    \gamma_a^b=\langle\Psi_{\text{NEO}}|a_a^b|\Psi_{\text{NEO}}\rangle
\end{equation}
and the protonic one-particle density matrix
\begin{equation}
    \label{eqn:protonic-1rdm}
    \gamma_A^B=\langle\Psi_{\text{NEO}}|a_A^B|\Psi_{\text{NEO}}\rangle\
\end{equation}
respectively, whereas their eigenvalues correspond to the occupation numbers. The FNOs with larger occupation numbers will contribute more to the total correlation energy, and those with a small occupation number will have insignificant contributions to the correlation energy and can therefore be discarded. Following this procedure, only the truncated set of unoccupied orbitals is used to define the active space and construct the pool of excitation operators in the ADAPT-VQE ansatz (Eq.~\eqref{eq:neo-adapt-ansatz}), thereby reducing the number of qubits and gates required for the quantum computation. Here, the electronic and protonic density matrices are computed with the NEO first-order Møller-Plesset (NEO-MP2)~\cite{pavosevic2020neooomp2,fetherolf2022multicomponent} wave function ($|\Psi_{\text{NEO}}\rangle$) as detailed in Sec. \ref{sec:theory:malonaldehyde}. For more detailed expressions of these, we refer to Eqns.~S56 and~S58 of Ref.~\citenum{pavosevic2020neooomp2}, respectively.

\subsection{Circuit compression with approximate quantum compiling}
\label{sec:isl}
The quantum circuits constructed via ADAPT-VQE, while compact relative to traditional approaches, may still exceed the limitation of current quantum hardware, particularly in two-qubit gate depth. To further reduce circuit depth, we apply ADAPT-AQC~\cite{jaderberg2025variational, adapt_aqc_github}.
In this variational approach, a parameterised circuit $\hat{V}(\vec{\theta})|0\rangle$ is optimised to approximate the target ADAPT-VQE state by minimizing the fidelity-based cost function
\begin{equation}
\label{eqn:aqc_cost}
    C = 1 - |\langle\Psi_{\text{ADAPT-VQE}}|\hat{V}(\vec{\theta})|0\rangle^{\bigotimes n}|^2.
\end{equation}
The variational ansatz is not assumed to have a fixed structure, but is instead built incrementally. At each increment, the algorithm adds a two-qubit unitary to the circuit, with the choice of qubits depending on the current state of optimization. For more details on the algorithm, we refer the reader to the literature~\cite{jaderberg2025variational, jaderberg2022solving}. Notably, this method is itself inspired by ADAPT-VQE, and the combination of both methods demonstrates the complexity of our approach. Furthermore, \acrshort{aqc} has been numerically demonstrated for up to 100 qubits and \acrshort{adapt-aqc} up to 50 qubits by using tensor network simulations~\cite{robertson2023approximate, jaderberg2025variational}. This indicates the potential scalability of the techniques in this work for helping to simulate proton transfer at large scales.

\subsection{Error mitigation using Zero-Noise Extrapolation}
\label{sec:err-mitigation}
The quantum computing \acrshort{neo} framework has been shown to produce accurate energy estimates in idealised, noise-free simulations~\cite{kovy2023jpcl,nyke2023}. In this work, we aim to study the effects of hardware noise and assess the ability to recover the expectation values through error mitigation. This involves performing simulations which take into account sampling (shot) noise and a realistic device noise model derived from noise characterization protocols on the \texttt{ibm\_fez} system\footnote[1]{The \texttt{ibm\_fez} quantum device contains 156 superconducting qubits on a Heron (2$^{\rm nd}$-generation) architecture. These qubits are connected via tunable-coupler technology in a heavy-hexagonal pattern.}. In this work, we apply the \acrshort{zne} technique~\cite{temme2017error, li2017efficient}, which involves deliberately amplifying the noise in a quantum circuit—typically by repeating certain gates—and measuring the resulting expectation values at different noise levels. By fitting these noisy results to an analytical curve and extrapolating back to the zero-noise limit, ZNE provides an improved estimate of the expectation value that would be obtained on an ideal, noise-free device. 

For noisy simulations, we use the device characterization data (see Appendix~\ref{sec:noise}), including qubit-specific $T_1$, $T_2$ coherence times, single- and two-qubit gate error rates, and readout errors, as determined from randomised benchmarking and echo protocols~\cite{magesan2012characterizing, krantz2019quantum}. The noise model selected corresponds to an Error Per Layered Gate (EPLG) of $0.003108$ across 18 qubits, representative of device behavior during the study.
The employed noise model includes (i) depolarizing channels derived by the 1- and 2-qubit gate error rate, followed by a thermal relaxation channel derived from the $T_1$, $T_2$ times and the corresponding gate lengths~\cite{qiskiterror}; 
(ii) bit-flip errors on the measurement outcomes derived from the readout error rates (see Appendix~\ref{sec:dev}).

Following the \acrshort{zne} description in~\cite{temme2017error}, we first run the original circuit to obtain the \textit{unmitigated} expectation value. Subsequently, copies of the circuit with folded gates to amplify the noise~\cite{majumdar2023best} are run, where the gate-folding steps are performed locally on randomly selected gates, using the Mitiq noise-mitigation software package~\cite{larose2022mitiq}. The noise-amplified circuits, alongside the original circuit, are executed using the Qiskit Aer simulation library~\cite{horii2023efficient} to compute noise-free and noisy expectation values. The resulting noisy expectation values are fit to a linear or quadratic function of the noise scaling factor $\lambda \in [1, 4]$, and extrapolated back to $\lambda \to 0$ to estimate noise-free energies and barrier heights, as demonstrated in Fig.~\ref{fig:zne-fit} and Fig.~\ref{fig:barrier-aqc}.

\section{Results and Discussion}
\label{sec:results}

\subsection{Barrier height evaluation}
\begin{figure*}
    \centering
    \includegraphics[width=\textwidth]{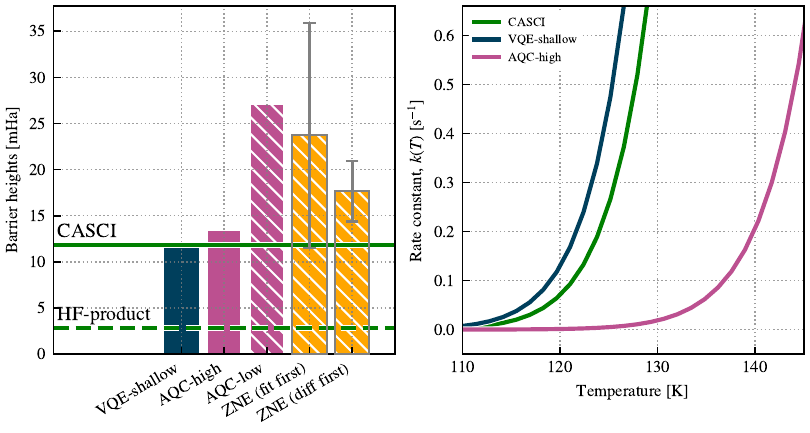}
    \caption{\textit{Left:} energy difference of the potential barrier (in mHa) for the VQE-shallow circuits, AQC-high and -low (noiseless, pink hatched), and the noisy AQC-low circuits with ZNE using two extrapolation methods (`fit first' and `diff first', yellow hatched). The horizontal solid line indicates the CASCI result, and the dashed line the HF-product result (without correlations) for the energy barrier, as a guideline. The ZNE-based bars indicate the median value sampled from 100 randomised gate-folded circuits (the first-to-third-quartile error bars span beyond the domain of the plot and are omitted for clarity). The AQC-based methods overestimate the barrier height in the noiseless regime and, on average, in the presence of noise. \textit{Right:} proton transfer rate constants, $k(T)$, as a function of temperature, $T$, computed using different methods based on using a quantum-corrected TST expression of Eq.~\eqref{eq:Eyring}. The green, blue, and pink lines indicate the CASCI, VQE-shallow, and AQC-high results, respectively. Energies estimated from AQC-low circuits would yield $k(T)\approx0$ in the temperature range shown, and are omitted for clarity.}
    \label{fig:rate_constants}
\end{figure*}

The computational pipeline introduced in Sec.~\ref{sec:methods} approximates the ground state wavefunction, with the aim of 
progressively reducing quantum circuit depth and thereby enable execution on near-term devices.
In Table~\ref{tab:summary-aqc} and Fig.~\ref{fig:rate_constants} (left panel), we report the computed energies and barrier heights, $\Delta E$, for proton transfer in malonaldehyde using various quantum and classical approaches, while the comparison of temperature dependence for the rate constants is shown in the right panel of Fig.~\ref{fig:rate_constants}. 
The CASCI result serves as a high-level quantum chemistry benchmark, while various quantum circuit-based methods demonstrate different trade-offs between circuit depth and fidelity. 
All quantum circuits presented in this work are transpiled to the \texttt{ibm\_fez} device with Heron r2 architecture~\cite{qpu_heron} and quantum resources requirements for their implementation are shown in Table~\ref{tab:summary-aqc}. 
\begin{table*}
    \centering
    \rowcolors{1}{white}{orange!20}
    \begin{tabular}{lllcccccc}
    \toprule
    Method & Label & State & 2Q-count & 2Q-depth  & Fidelity & $E$ [mHa] & $\Delta E$ [mHa] \\ 
    \midrule
    FNO-NEO-CASCI           & CASCI & \textit{Left}   & --   & --  & --    & -600.666        \\
                            &  & \textit{Middle} & --   & --  & --    & -588.809 & 11.857 \\
    HF-product           & HF-product & \textit{Left}   & 0    & 0   & 0.888    & -552.090        \\
                            &  & \textit{Middle} & 0    & 0   & 0.936    & -549.241 & 2.850 \\
    ADAPT-VQE (deep)        & VQE-deep & \textit{Left}   & 1844  & 1362 & 0.998    & -599.268        \\
                            &  & \textit{Middle} & 940  & 662 & 0.999    & -587.497 & 11.771 \\
    ADAPT-VQE (shallow)     & VQE-shallow & \textit{Left}   & 551  & 411 & 0.971    & -591.237        \\
                            &  & \textit{Middle} & 271  & 211 & 0.976    & -579.609 & 11.628 \\
    ADAPT-AQC (high)        & AQC-high & \textit{Left}   & 81   & 51  & 0.961  & -578.785        \\
                            &  & \textit{Middle} & 405  & 90  & 0.967  & -565.358 & 13.427 \\
    ADAPT-AQC (low)         & AQC-low & \textit{Left}   & 81   & 51  & 0.961  & -578.785        \\
                            &  & \textit{Middle} & 85   & 12  & 0.953 & -551.645 & 27.140 \\
    ADAPT-AQC (low+ZNE)     & ZNE (fit first) & \textit{Left}   & --   & --  & --  & $-548\pm7$       \\ 
                            &  & \textit{Middle} & --   & --  & -- & $-524\pm10$ & $24\pm12$ \\ 
                            &  ZNE (diff first) & \textit{Left}, \textit{Middle} & --   & --   & --  & --  & $18\pm3$        \\
    \bottomrule
    \end{tabular}
    \caption{
     Summary of ADAPT-VQE and ADAPT-AQC circuits, along with energies and proton transfer barriers computed via statevector simulation and classical methods. Two-qubit gate depths and counts refer to circuits transpiled for IBM Heron processors. Circuit fidelities are given with respect to the FNO-NEO-CASCI wavefunction. To improve readability and focus on meaningful energy differences, a constant offset of 265 Ha was added to reported absolute energies.
    }
\label{tab:summary-aqc}
\end{table*}

The Hartree-Fock state significantly underestimates $\Delta E$ compared to CASCI reference. This highlights the critical role of correlation for accurate barrier evaluation and the need for quantum circuits balancing entanglement depth with hardware feasibility. For this, we apply two layers of approximations as discussed in Sec.~\ref{sec:methods}. First, employing ADAPT-VQE in noiseless statevector simulation regime, we construct circuits labelled VQE-deep and VQE-shallow yielding energies within $10^{-3} \textrm{ Ha}$ and $10^{-2} \textrm{ Ha}$ of the CASCI ground state, respectively (see Table~\ref{tab:summary-aqc}). The resource requirements of VQE-deep circuits are too large for execution on the currently available quantum processors, while it delivers outstanding accuracy for energy barrier and absolute energies. Although VQE-shallow produces circuits with relatively less overhead and delivers outstanding accuracy with respect to CASCI, their practical deployment remains unrealistic in the near term. Specifically, the two-qubit gate depth is the most important for reducing noise on current superconducting quantum computers, where decoherence over time is more significant than errors from imperfect execution of gates. 

We further optimise the VQE-shallow circuits using ADAPT-AQC running with two settings. The first, ADAPT-AQC (high), is set to retain high fidelity relative to the target circuit at the cost of a deeper solution. The second, ADAPT-AQC (low), is set to prioritise reducing the entangling gate depth, at the cost of lower fidelity. Running the algorithm under both of these settings produces two sets of circuits we label AQC-high and AQC-low, as shown in Table~\ref{tab:summary-aqc}.  

The AQC-high circuits are able to reduce the two-qubit gate depth of the VQE-shallow circuits by $88\%$ and $57\%$ for the \textit{Left} and \textit{Middle} systems, respectively, while still maintaining fidelity $99\%$. Furthermore, evaluating the states produced by the AQC-high circuits yields a barrier height within approximately 10\% of the CASCI reference, but leads to a 60\% underestimate in the rate constant at 120 K (Fig.~\ref{fig:rate_constants}, right panel). These discrepancy arises due to the exponential dependence of rate constants on barrier height.
To quantify the sensitivity of the rate constant to barrier height errors, we note that an error $\delta E$  in the barrier leads to a fractional error in the rate constant $\delta k/k $ approximately given by $\delta k/k \approx -\delta E / k_BT $. At 120 K, this implies that the barrier must be accurate to within $\approx$ 0.08 mHa ($\approx$ 2 meV) in order to keep the error in the corresponding rate constant below 20\%. This sets a stringent requirement on quantum energy estimation algorithms to yield reliable kinetics in this temperature regime. Reaching this level of precision remains a key challenge—and benchmark—for emerging quantum algorithms applied to chemical reaction dynamics. 

The AQC-low circuits have a minimum fidelity of $97\%$ with respect to the VQE-shallow target. Whilst no improvement over the AQC-high could be achieved for the \textit{Left} system, a solution for the \textit{Middle} circuit is found with only 12 two-qubit gate depth. Thus, the AQC-low circuits have small enough resource requirements to be realised on currently available quantum hardware.
However, the cost of the significant depth reduction of the ADAPT-AQC (low) protocol means the evaluated circuits overestimate the barrier by approximately 60\%, resulting in a drastic suppression of the computed rate constant by more than two orders of magnitude within the same temperature range. This result underscores the importance of high-precision energy estimation when using quantum algorithms for chemical kinetics. Nevertheless, we accept these inaccuracies as a trade-off in order to evaluate the feasibility of using the NISQ devices, which remain limited by the depth of circuits they can reliably execute.

Our results show that noiseless simulations can recover the proton transfer energy barrier accurately with significant compression from \gls{adapt-aqc}. Circuits produced by AQC-high significantly reduce the depth of entangling gates compared to VQE-shallow while maintaining good accuracy; however, they are still too deep to be executed reliably on noisy devices. Further compromising the fidelity in favor of lower depth, the circuits produced by AQC-low offer a more realistic scenario for noisy quantum hardware, though they severely overestimate the barrier. Nonetheless, this may represent the only currently feasible option for application on a real device.

\subsection{Proton transfer dynamics}
To simulate the proton transfer processes, we construct 7 Hamiltonians defined along the adiabatic trajectory by interpolating between \textit{Left}, \textit{Middle}, and \textit{Right} as defined in Eq.~\eqref{eq:HLMRT}.
We use the nomenclature LMR to label the resulting time-dependent Hamiltonians and corresponding states, where each digit reflects the weight of the respective \textit{Left}, \textit{Middle}, and \textit{Right} setups. For example, the label 210 corresponds to the following Hamiltonian 
\begin{equation}\label{eq:HLMR}
\hat{H}(210) = \frac{2}{3}\hat{H}_{\rm Left} + \frac{1}{3}\hat{H}_{\rm Middle} + 0 \hat{H}_{\rm Right}. 
\end{equation}
We then repeat the application of our pipeline in Fig.~\ref{fig:workflow} on each of these Hamiltonians to produce a ground state for each intermediate point that can be run on current quantum hardware. Specifically, we use ADAPT-VQE (shallow), followed by ADAPT-AQC (low) to produce 7 circuits that approximate the ground state wavefunctions along the adiabatic proton transfer process for the time dependent Hamiltonian in Eq.~\eqref{eq:HLMRT}.  
Fig.~\ref{fig:barrier-aqc} illustrates the full adiabatic proton‐transfer pathway from the \textit{Left} to the \textit{Right} states. The plotted energies correspond to expectation values of the NEO Hamiltonian without the addition of the scalar constant arising from classical nuclear and frozen-core electron interactions. The hardware requirements for these circuits, along with the corresponding errors, are summarised in Table~\ref{tab:aqc-circuits}. One can observe shifts in energy across all states, which are not always systematic. In addition, asymmetries in the two-qubit gate depth of circuits are evident. Both effects originate from asymmetries in the ADAPT-VQE procedure, which initially uses the Hartree--Fock state of the \textit{Left} setup (see Figure~\ref{fig:TOC}) as a starting point for all steps along the adiabatic trajectory. These initial imbalances are inherited and further amplified by ADAPT-AQC compression.

\begin{table}
   \small
   \centering
   \begin{tabular}{lcccc}
   \toprule
   \textbf{State} & \textbf{Error (Ha)} & \textbf{2Q-depth} & \textbf{2Q-count} \\ 
   \midrule
    300 (\textit{Left})  & 0.022 &  51 & 81 \\
    210 & 0.043 & 12 & 87 \\
    120 & 0.037 & 12 & 86 \\
    030 (\textit{Middle}) & 0.037 & 12 & 85 \\
    021 & 0.039 & 12 & 87 \\
    012 & 0.044 & 20 & 67 \\
    003 (\textit{Right}) & 0.030 & 16 & 130 \\

   \bottomrule
   \end{tabular}
   \caption{Summary of circuits contracted with ADAPT-AQC low. \textit{Left} and \textit{Middle} are the same as in Table \ref{tab:summary-aqc}. The energy errors, in Hartree, are relative to the CASCI energy; we also report the two-qubit depth and the number of two-qubit gates in each synthesised circuit after transpiling for the \texttt{ibm\_fez} heavy-hexagonal topology.}
   \label{tab:aqc-circuits}
\end{table}

\begin{figure*}
    \centering
    \includegraphics[width=\textwidth]{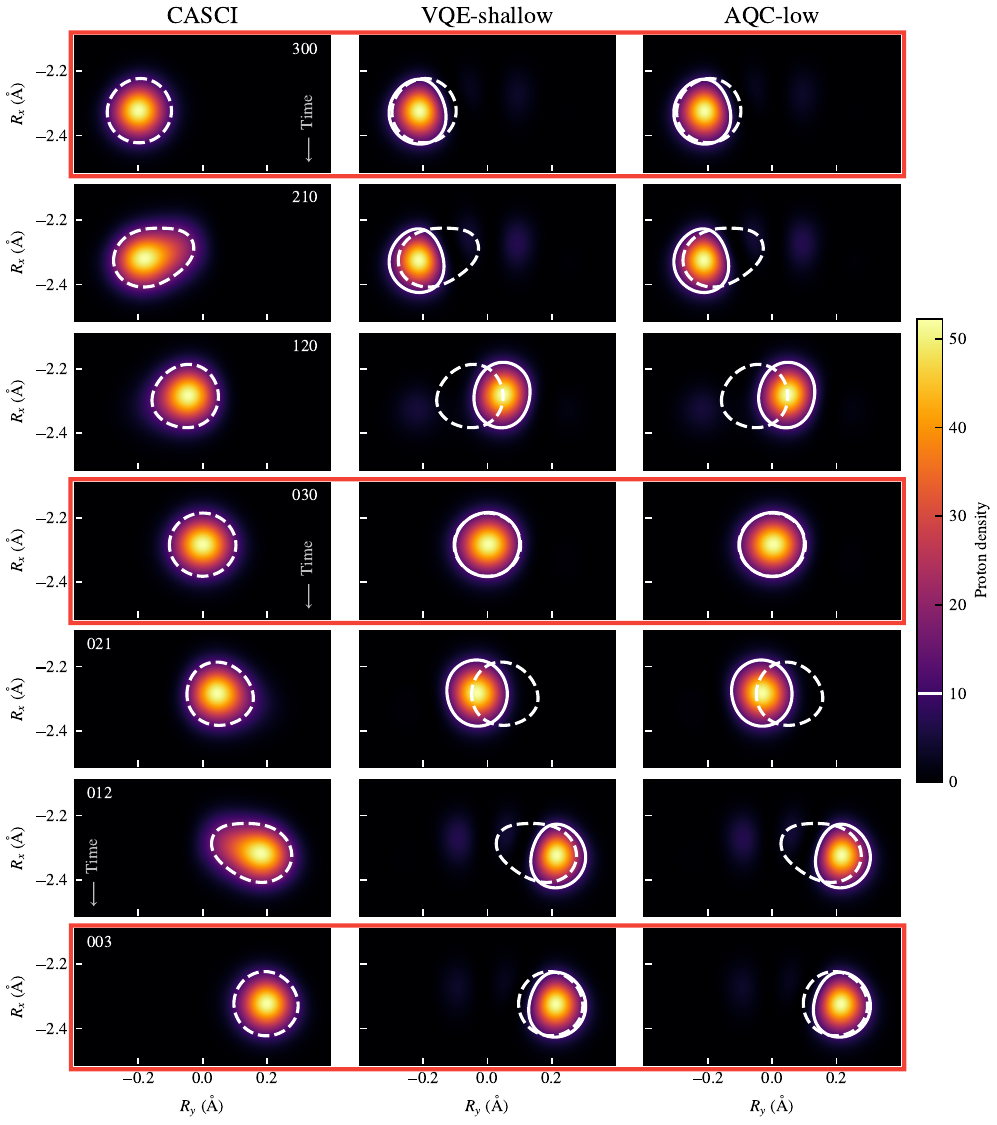}
        \caption{
        Proton density distributions obtained using the CASCI, VQE-shallow, and AQC-low methods. Each subplot corresponds to one computational method and shows the proton density for 7 positional configurations from \textit{Left} (300) to \textit{Right} (003) states. The \textit{Left}, \textit{Middle}, and \textit{Right} densities (red boxes) most accurately reproduce the CASCI references, and their corresponding energies are used in barrier evaluation. The densities are evaluated on a 2D grid in the XY plane located at $R_z = 0.0$~\AA.
    }
    \label{fig:proton_density}
\end{figure*}

For illustration purposes, we plot the proton densities along the proton transfer process in malonaldehyde. Figure~\ref{fig:proton_density} compares these proton densities as computed using the increasing approximations of our pipeline: a high-accuracy CASCI reference (left), the VQE-shallow solution (middle), and the AQC-low solution (right).

The CASCI calculations show the exact nuclear delocalization at each geometry along the adiabatic transfer trajectory. The VQE-shallow solution qualitatively reproduces the key features of the nuclear density, particularly capturing the delocalization and correct spread of the proton wave packet at the \textit{Left}, \textit{Middle}, and \textit{Right} states. However, at intermediate points along the reaction path, the VQE-shallow wave packets appear somewhat more localised than in the CASCI reference, suggesting that the variationally constructed ansatz, though compact, does not fully capture all the correlations needed for accurate delocalization at these states.

This limitation stems from the core approximations in ADAPT-VQE. The ansatz is constructed iteratively from a predefined operator pool, targeting ground-state energy minimization at each step. While this approach ensures efficiency and adaptivity, it can under-represent entanglement or spatial correlation effects that are not sufficiently prioritised by the energy gradient criterion used in operator selection, and these inaccuracy persist even in VQE-deep.

The right panel in Figure~\ref{fig:proton_density} demonstrates how, despite its reduced depth, AQC-low solution almost exactly reproduces the key features of the VQE-shallow densities across the entire transfer pathway, including the subtle asymmetries and spatial localization patterns. This highlights the utility of \gls{aqc} not only as a tool for circuit compression but also as a robust method for preserving chemically and physically meaningful wavefunction features in practical quantum simulations.

\subsection{Hardware noise simulations}
\begin{figure*}
    \centering
    \includegraphics[width=\linewidth]{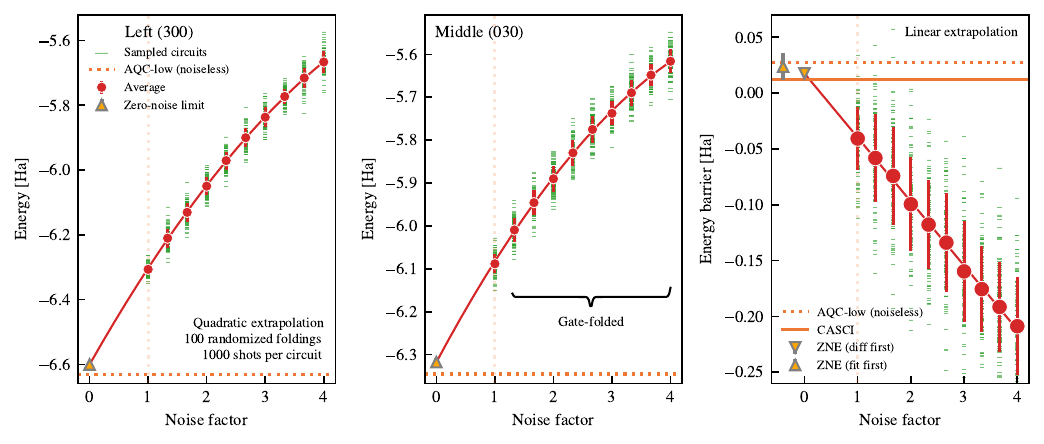}
    \caption{Zero-noise extrapolation (ZNE) results for energy expectation values and proton transfer barrier heights across representative proton configurations in malonaldehyde, obtained from AQC-low circuits simulated under \texttt{ibm\_fez} noise models. Ground-state energy expectation value (in Ha) as a function of noise amplification factor for \textit{Left} (300, left panel) and \textit{Middle} (030, central panel) proton configurations; the zero-noise limit (yellow markers) is obtained by extrapolation using a quadratic fit. The corresponding barrier energy, computed as the difference of the extrapolated intercepts, is indicated in the right panel as `Fit first' (upward-pointing yellow marker). On the right, we also show the barrier height measurements computed by the mean difference in raw energy between the \textit{Middle} and \textit{Left} states for each noise amplification factor (red markers). The zero-noise limit (downward-pointing yellow marker, `Difference first') is computed via a linear fit to the data (red line). Both `Fit first' and `Difference first' results are obtained by fitting the mean expectation values; the error bars are the variance in the correlation matrix element corresponding to the fitted intercept. In all panels, we indicate the noiseless energy value and barrier height from the AQC-low circuits (dotted lines); in the right panel, we indicate the CASCI reference energy barrier (solid horizontal line).}
    \label{fig:zne-fit}
\end{figure*}

\begin{figure*}
    \centering
    \includegraphics[width=\linewidth]{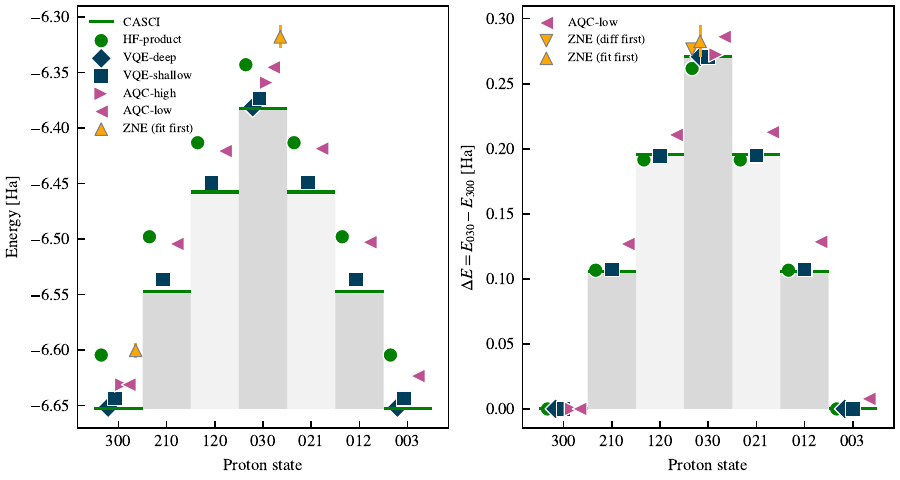}
    \caption{Energy profiles and relative barrier heights for the {Left}, {Middle}, {Right} and intermediate states of proton transfer in malonaldehyde computed along the adiabatic trajectory, comparing the following simulation methods: CASCI (green lines), HF-product (green circles), VQE-deep (blue diamonds), VQE-shallow (blue squares), AQC-high (purple right-pointing triangles), AQC-low (purple left-pointing triangles), ZNE fit-first (yellow upward-pointing triangles). \textit{Left:} Absolute ground-state energies (in Ha) as a function of discrete proton configurations (denoted by the three-digit labels \texttt{300}$\to$\texttt{210}$\to$\texttt{120}$\to$\texttt{030}$\to$\texttt{021}$\to$\texttt{012}$\to$\texttt{003}). \textit{Right:} Energy barriers relative to the Left state energy, $\Delta E = E_{\mathrm{state}} - E_{300}$ (in Ha). The circuits compiled with AQC-high achieve excellent agreement with the VQE and CASCI values.}
  \label{fig:barrier-aqc}
\end{figure*}

In the following, we present results from the noisy simulations of the \textit{Left} and \textit{Middle} circuits, transpiled for the \texttt{ibm\_fez} backend and executed with ZNE error mitigation. 
The ground‐state energies for the \textit{Left} (300) and \textit{Middle} (030) configurations obtained with the AQC-low circuits under the \texttt{ibm\_fez} noise model (see Table~\ref{tab:timestamps}) are evaluated for noise amplification factor $\lambda \in [1, 4]$. For each value of $\lambda$, we constructed 100 distinct circuits with randomised folding of the 2-qubit gates. Each noise‐scaled circuit is sampled 1000 times, and the noisy expectation values are then fitted to a quadratic function (See Appendix \ref{sec:model_selection_bootstrap} for a discussion about model selection). The difference between the estimated intercepts for the \textit{Left} and \textit{Middle} configurations is used as an estimate for the ZNE energy. We refer to this as the `fit first' method. The error bars in the ZNE energy estimates are given by standard error of the corresponding intercepts, which are added in quadrature to yield the $1\sigma$ uncertainty of the `fit first' barrier energy. In Figure~\ref{fig:zne-fit} we plot the measured energies for the $H_{030}$ and $H_{300}$ along with the fitted lines that are used to estimate the ZNE energy of the barrier. 
Notably, the ZNE correction applied to the noisy simulations fails to recover the absolute energies for \textit{Left} and \textit{Middle} states, landing them slightly above HF results (see Table~\ref{tab:summary-aqc} and Figure~\ref{fig:barrier-aqc}). Thus, the simulations incorporating realistic quantum hardware noise cannot capture the correlations effects under error mitigation conditions with ZNE. However, the barrier computed from these zero‐noise limit (`fit first', yellow upward triangles in Figure~\ref{fig:zne-fit}) closely approaches the ideal, noiseless circuit energies (dotted lines in the left panel of Figure~\ref{fig:zne-fit}). While this agreement may be partly coincidental --- given that ZNE fails to recover the absolute energies --- it suggests that the noise-scaling fit may better capture energy differences than absolute values.  

Motivated by the above observation, we explore an alternative approach in which the energy difference between \textit{Middle} and \textit{Left} is first computed at each $\lambda$, and then the difference is extrapolated to $\lambda=0$. We refer to this method as `difference first'. In the right panel of Figure~\ref{fig:zne-fit} we plot this energy difference along with a fitted, linear function (see Appendix \ref{sec:model_selection_bootstrap} for a discussion about model selection). 
The `difference first' method yields a barrier height of $18\pm3$ mHa (yellow, downward triangles in Figure~\ref{fig:barrier-aqc}), which is lower than the `fit first' result ($24\pm12$ mHa). The uncertainty is also smaller (which is confirmed using non-parametric bootstrap estimation, see Appendix \ref{sec:model_selection_bootstrap} for details), placing the noiseless AQC-low results $2\sigma$ away from the so-obtained zero-noise limit. Further investigation is needed to determine whether this effect is systematic or accidental.

These results suggest that, despite aggressive circuit compression the proposed framework remains too resource-intensive for near-term devices and will likely only become practical in the FTQC regime.  

\section{Conclusions}

This work establishes a practical framework for simulating proton transfer with quantum resources that are compatible with both near-term and future quantum computing architectures. By combining nuclear-electronic orbital methods with adaptive quantum circuit construction (ADAPT-VQE and ADAPT-AQC), we demonstrate that it is possible to approximate the proton transfer barrier in malonaldehyde with high accuracy while significantly reducing quantum circuit depth. Notably, our shallowest circuits achieve chemical insight with only 81 two-qubit gates, positioning them near the frontier of feasibility for current hardware, while deeper circuits retain higher fidelity to reference energies and point the way toward early fault-tolerant implementations.

Beyond malonaldehyde, this methodology offers a modular pipeline for tackling a broader class of proton-coupled electron transfer problems, where the interplay of light nuclei and electronic structure challenges classical methods. As quantum hardware continues to improve, the techniques presented here offer a path toward achieving chemical accuracy in simulating proton dynamics, representing a promising early application for the first generation of fault-tolerant quantum computers.

\begin{acknowledgments}
    This research was supported by funding from the Wallenberg Center for Quantum Technology (WACQT), and the NCCR MARVEL, a National Centre of Competence in Research, funded by the Swiss National Science Foundation (grant number 205602).
    This work was supported by the Hartree National Centre for Digital Innovation, a UK Government-funded collaboration between STFC and IBM.
    For the purpose of open access, the authors have applied a Creative Commons Attribution (CC BY) license to any Author Accepted Manuscript version arising from this submission.
    IBM, the IBM logo, and \href{https://ibm.com}{www.ibm.com} are trademarks of International Business Machines Corp., registered in many jurisdictions worldwide. Other product and service names might be trademarks of IBM or other companies. The current list of IBM trademarks is available at \href{https://www.ibm.com/legal/copytrade}{www.ibm.com/legal/copytrade}.

    The research in this paper made use of the following software packages and libraries: 
    \textsc{Python}~\cite{van1995python},
    \textsc{Numpy}~\cite{harris2020array},
    \textsc{Scipy}~\cite{virtanen2020scipy},
    \textsc{Matplotlib}~\cite{hunter2007matplotlib, caswell2020matplotlib},
    \textsc{Unyt}~\cite{goldbaum2018unyt},
    \textsc{Qiskit}~\cite{javadi2024quantum}.
    
\end{acknowledgments}

\bibliography{main}
\clearpage

\appendix

\section{Basis set setups}\label{sec:basis}
The protonic orbitals are constructed using the \gls{pb4}~\cite{yuqi2020} basis set, which is significantly larger than the \gls{dzsnb}~\cite{webb2002} used in~\citer{kovy2023jpcl}. For the electronic orbitals centered at positions of proton transfer, we use the ghost basis functions from the \gls{cc5}~\cite{dunn1989} basis 
set instead of \gls{631}. The \gls{sto6}~\cite{STO6G}, 
\gls{631}~\cite{631G}, and \gls{ccd}~\cite{dunn1989} basis sets are used for the peripheral hydrogen, carbon, and oxygen atoms, 
respectively. This choice of basis sets improved the physical accuracy of the orbitals involved in \gls{pt}, leading to a more realistic representation of the underlying physics, while maintaining the total number of orbitals tractable.

\section{Cartesian coordinates and proton positions for malonaldehyde setups}\label{sec:xyz}

Table~\ref{tab:xyz_all} reports the full set of Cartesian coordinates (in Bohr) for the nine‐atom malonaldehyde system in three representative proton‐transfer configurations: the \textit{Left}, \textit{Middle}, and \textit{Right} setups. Each configuration involves two oxygen atoms, three carbon atoms, and four hydrogen atoms (one of which participates directly in the intramolecular hydrogen bond). The molecular origin is placed at the central carbon atom, and all atomic positions are given relative to this point. The \textit{Left} and Right geometries correspond to the proton localised on each oxygen, whereas the \textit{Middle} geometry represents the proton symmetrically shared between the two oxygens.

\begin{table}[htbp]
  \centering
    \caption{Cartesian coordinates (in \AA) for all nine atoms in the Left, Middle, and Right malonaldehyde setups.}
  \label{tab:xyz_all}
    \small
    \begin{tabular}{@{}lrrr@{}}
      \toprule
        & $x$ & $y$ & $z$ \\
      \addlinespace
      \multicolumn{4}{@{}l@{}}{\textit{Left} setup}\\
      \midrule
      O & -2.01516150 &  1.25951633 &  0.00000000 \\
      O & -2.05554525 & -1.16937003 &  0.00000000 \\
      C & -0.74721061 &  1.23462704 &  0.00000000 \\
      C & -0.72001126 & -1.17773733 &  0.00000000 \\
      C &  0.00000000 &  0.00000000 &  0.00000000 \\
      H &  1.08331158 &  0.00000000 &  0.00000000 \\
      H & -0.20901142 &  2.19223326 &  0.00000000 \\
      H & -0.26674493 & -2.16891499 &  0.00000000 \\
      H & -2.22176148 & -0.18711600 &  0.00000000 \\
      \addlinespace
      \multicolumn{4}{@{}l@{}}{\textit{Middle} setup}\\
      \midrule
      O & -2.04864177 &  1.15865439 &  0.00000000 \\
      O & -2.04864177 & -1.15865439 &  0.00000000 \\
      C & -0.74990078 &  1.19199519 &  0.00000000 \\
      C & -0.74990078 & -1.19199519 &  0.00000000 \\
      C &  0.00000000 &  0.00000000 &  0.00000000 \\
      H &  1.08244928 &  0.00000000 &  0.00000000 \\
      H & -0.27965698 &  2.17995088 &  0.00000000 \\
      H & -0.27965698 & -2.17995088 &  0.00000000 \\
      H & -2.18129112 &  0.00000000 &  0.00000000 \\
      \addlinespace
      \multicolumn{4}{@{}l@{}}{\textit{Right} setup}\\
      \midrule
      O & -2.05554525 &  1.16937003 &  0.00000000 \\
      O & -2.01516150 & -1.25951633 &  0.00000000 \\
      C & -0.72001126 &  1.17773733 &  0.00000000 \\
      C & -0.74721061 & -1.23462704 &  0.00000000 \\
      C &  0.00000000 &  0.00000000 &  0.00000000 \\
      H &  1.08331158 &  0.00000000 &  0.00000000 \\
      H & -0.26674493 &  2.16891499 &  0.00000000 \\
      H & -0.20901142 & -2.19223326 &  0.00000000 \\
      H & -2.22176148 &  0.18711600 &  0.00000000 \\
      \bottomrule
    \end{tabular}
\end{table}

\begin{table}
    \caption{Expectation values of position operator (in \AA),
    obtained from various ans\"atze. ``Orb. center'' referes to the position of the orbital center revealed from MP2 calculation.
    }
    \begin{tabular}{lrrr}
        \toprule
        \textbf{Method} & $x$ & $y$ & $z$ \\
        \midrule
        \multicolumn{4}{c}{\textit{Left} setup --- 300} \\
        Orb. center & -2.22176148 & -0.18711600 & 0.00000000 \\
        CASCI       & -2.32501550 & -0.19558330 & 0.00000000 \\
        ADAPT       & -2.32567623 & -0.19555274 & 0.00000000 \\
        AQC         & -2.32567623 & -0.19555274 & 0.00000000 \\
        \hline
        \multicolumn{4}{c}{\textit{Middle} setup --- 030} \\
        Orb. center & -2.18129112 & 0.00000000 & 0.00000000 \\
        CASCI       & -2.28629872 & 0.00000000 & 0.00000000 \\
        ADAPT       & -2.28548356 & 0.00007774 & 0.00000000 \\
        AQC         & -2.28548356 & 0.00007774 & 0.00000000 \\
        \bottomrule
    \end{tabular}
    \label{tab:pos}
\end{table}

\section{Characteristics of IBM Quantum devices}
\label{sec:dev}
\label{sec:noise}

The \texttt{ibm\_fez} device is a superconducting quantum processor built on IBM's second-generation Heron architecture, implementing 156 fixed-frequency transmon qubits arranged in a heavy-hexagonal lattice topology. This design minimises qubit crosstalk and enables efficient connectivity through tunable coupler technology~\cite{krantz2019quantum,ibm_heron_docs}. Each qubit is characterised by individual relaxation ($T_1$), dephasing ($T_2$), single- and two-qubit gate fidelities, and readout error rates, which are determined via standard quantum benchmarking protocols, including randomised benchmarking and Hahn echo sequences~\cite{magesan2012randomized,krantz2019quantum}. Device performance is monitored using the Error Per Layered Gate (EPLG) metric~\cite{mckay2023benchmarking}, which quantifies the cumulative noise contribution across layers of a quantum circuit. A summary of noise models sampled is shown in Table~\ref{tab:timestamps}. For this work, we selected a representative noise model based on calibration data recorded on November, 10th 2024 at 10:40:31 UTC, which yielded the lowest 18-qubit EPLG of 0.003108 across the sampled noise models and had the smallest difference between the calibration and sampling times. The corresponding analytic noise model incorporates depolarizing gate errors, thermal relaxation channels, and qubit-specific readout errors as captured by IBM's Qiskit Aer simulator. While these characteristics represent the current state-of-the-art in superconducting quantum hardware, they continue to limit the accuracy of quantum simulations requiring high energy resolution, such as those involved in proton transfer dynamics.

\begin{figure*}
    \centering
    \includegraphics[width=\linewidth]{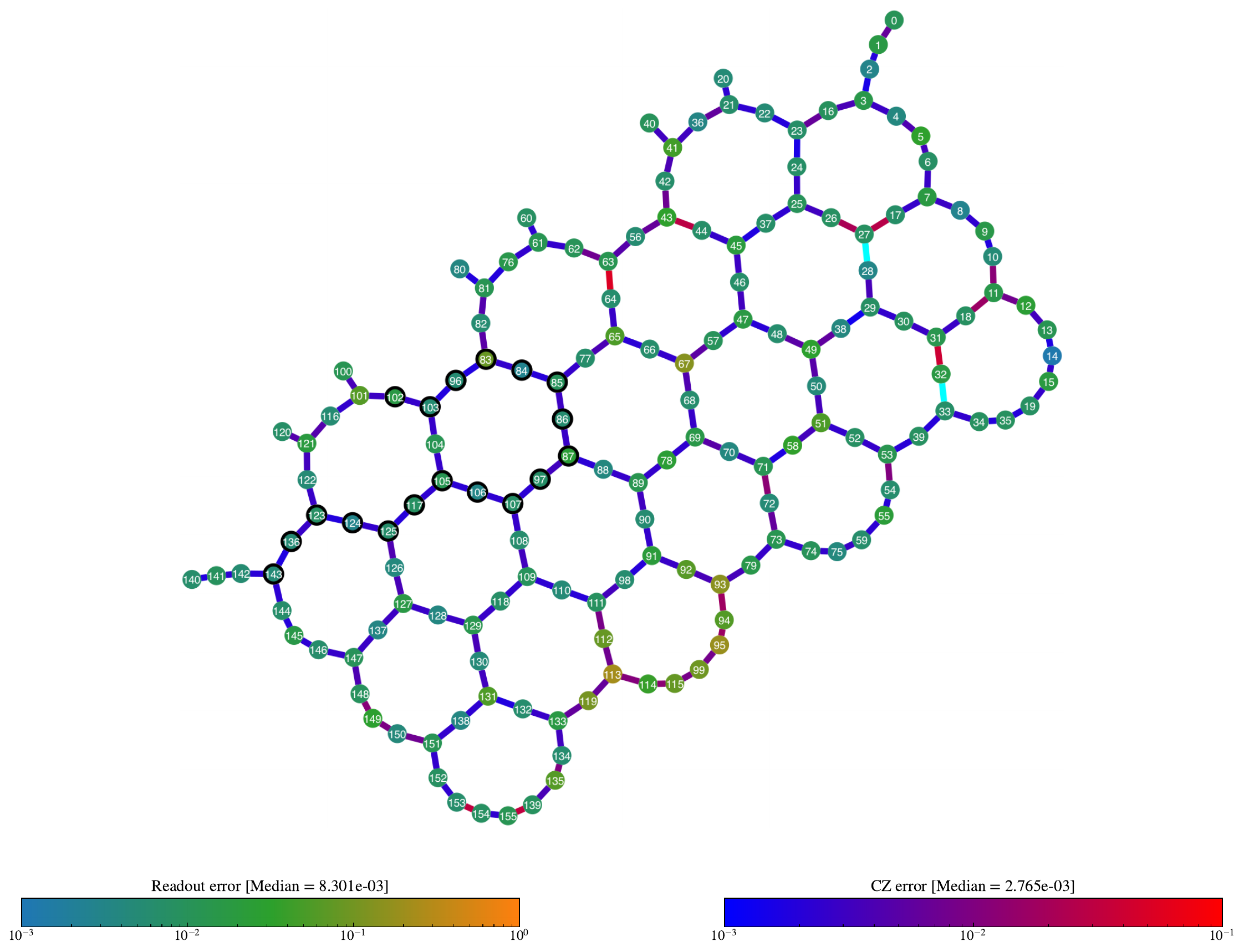}
    \caption{Error map of the \texttt{ibm\_fez} processor captured for the calibration data recorded on 10 November 2024 at 10:40:31 UTC. Node colour represents the readout error of the physical qubits, and the edge colour represents the CZ error. Calibration data for two connections were not available (denoted in cyan). The physical qubits that correspond to a representative transpiled AQC-low circuit for the 030 state with a noise scale factor of 3 are indicated by black circles.}
    \label{fig:enter-label}
\end{figure*}

\begin{table}[htbp]
    \caption{Date-time stamps for the noise models, the date-time the device was last calibrated, and error per layered gate (EPLG) for 18 qubits. The noise model selected for simulations is highlighted.}
    \centering
    \begin{tabular}{c@{\hspace{0.5cm}}cc}
        \toprule
        \textbf{Date-time (UTC)} & \textbf{Last update (UTC)} & EPLG-18\\
        \midrule
        2024-09-30 21:18:12 & 2024-09-30 20:50:06& 0.003253 \\
        2024-10-01 02:44:40 & 2024-10-01 02:03:12& 0.003376 \\
        2024-10-01 09:50:30 & 2024-10-01 08:59:31& 0.003376 \\
        2024-10-01 15:50:16 & 2024-10-01 14:47:47& 0.003376 \\
        2024-11-09 21:47:33 & 2024-11-09 20:34:16& 0.003163 \\
        2024-11-10 04:41:32 & 2024-11-10 02:34:50& 0.003108 \\
\rowcolor{orange!20}   \textbf{2024-11-10 10:40:31} & \textbf{2024-11-10 10:29:01}& \textbf{0.003108} \\
        2024-11-10 16:41:08 & 2024-11-10 15:35:14& 0.003108 \\
        2024-11-30 22:31:06 & 2024-11-30 21:45:55& 0.003515 \\
        2024-12-01 04:37:21 & 2024-12-01 03:57:18& 0.003515 \\
        2024-12-01 10:58:53 & 2024-12-01 10:05:27& 0.003515 \\
        2024-12-01 16:54:28 & 2024-12-01 16:31:06& 0.003515 \\
        2024-12-31 22:56:05 & 2024-12-31 18:32:04& 0.003515 \\
        2025-01-01 01:23:56 & 2025-01-01 00:54:59& 0.003515 \\
        2025-01-01 09:52:17 & 2025-01-01 01:23:56& 0.003515 \\
        2025-01-31 22:54:03 & 2025-01-31 22:07:43& 0.003858 \\
        2025-02-01 04:41:04 & 2025-02-01 04:14:31& 0.003490 \\
        2025-02-01 10:33:22 & 2025-02-01 10:20:50& 0.003490 \\
        2025-02-01 15:38:12 & 2025-02-01 14:06:44& 0.003490 \\
        \bottomrule
    \end{tabular}
    \label{tab:timestamps}
\end{table}

\section{Zero-noise extrapolation with bootstrap resampling}
\label{sec:model_selection_bootstrap}
\begin{figure*}
    \centering
    \includegraphics[width=\linewidth]{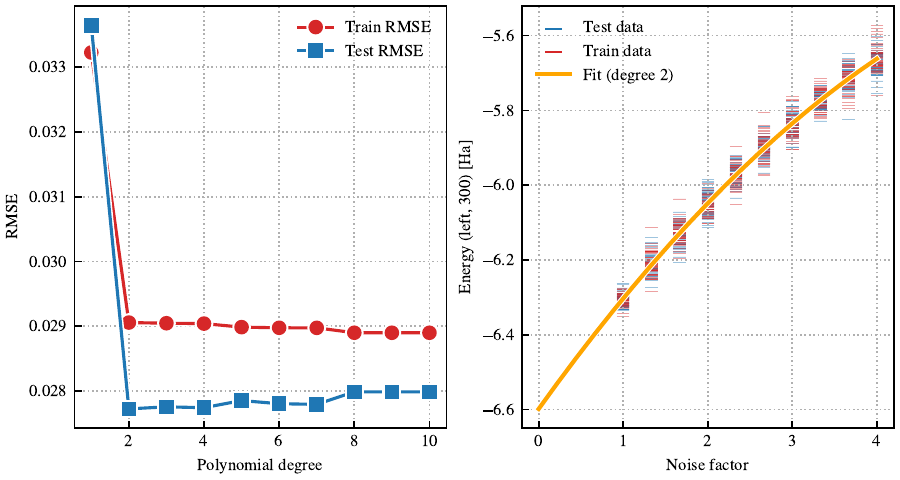}
    \caption{RMSE as a function of polynomial degree (left panel) and optimal fit to data (right panel) for the energies of the left system $H_{300}$.}
    \label{fig:polyfit_300}
\end{figure*}

\begin{figure*}
    \centering
    \includegraphics[width=\linewidth]{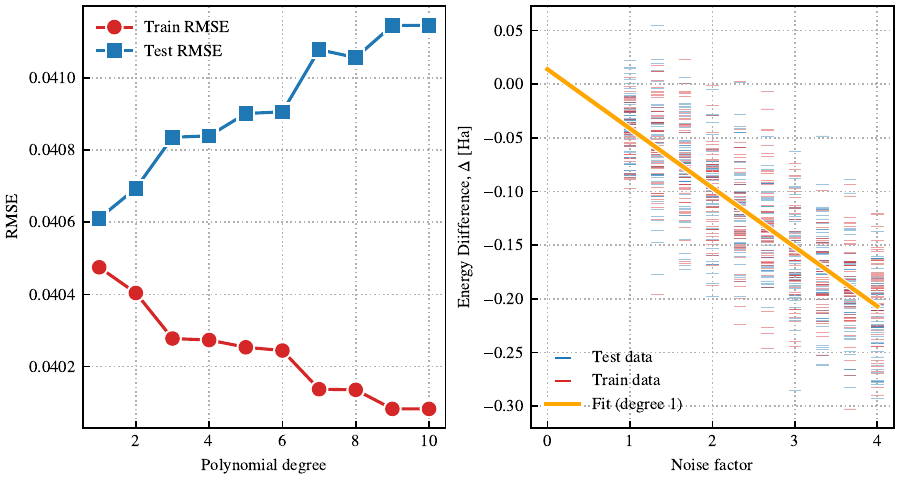}
    \caption{RMSE as a function of polynomial degree (left panel) and optimal fit to data (right panel) for the difference in energies between the middle and left systems $\Delta = E_{030} - E_{300}$.}
    \label{fig:polyfit_diff}
\end{figure*}

To identify the most appropriate model describing the relationship between the noise factor and the measured energy output from the quantum device, polynomial regression models of varying degrees were evaluated. Model performance was quantified using the root mean squared error (RMSE), calculated separately for a hold out test set where the optimal polynomial degree was determined as the one minimizing this quantity, thereby achieving a balance between model complexity and generalizability while avoiding bias towards the intercept at $\lambda = 0$. In Figure~\ref{fig:polyfit_300}, the RMSE is shown as a function of polynomial degree for the measured energies of the left system ($H_{300}$). As illustrated, the lowest RMSE is attained for a polynomial degree of $2$.
In Figure~\ref{fig:polyfit_diff}, the same procedure is applied to the difference in energies between the middle and left systems $\Delta E = E_{030} - E_{300}$ which is used to compute the `difference first' barrier height shown in Tab.~\ref{tab:summary-aqc}. In this case, the optimal polynomial degree is found to be $1$. 
In both figures, model fits corresponding to the selected optimal polynomial degrees are presented alongside the observed data to enable qualitative assessment of model performance. 

\begin{figure*}
    \centering
    \includegraphics[width=\linewidth]{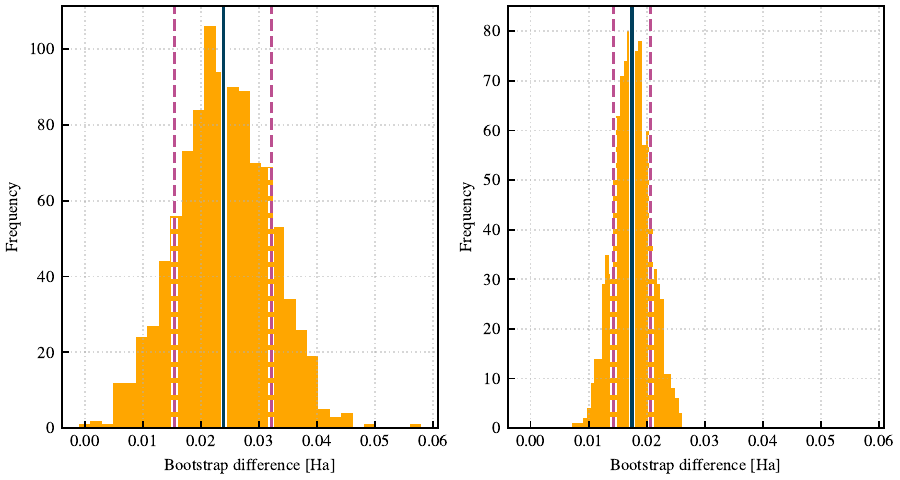}
    \caption{Empirical bootstrap distributions for the intercept for the `Fit first' (left panel) and `Diff first' (right panel) extrapolation methods. The solid line shows the median and dashed lines indicate the $15^{\rm th}$ and $85^{\rm th}$ percentiles, respectively.}
    \label{fig:bootstrap_intercepts}
\end{figure*}

Given the optimal polynomial degree, bootstrap resampling was employed to estimate the variability and confidence intervals for the intercept parameter of the regression model corresponding to the error-mitigated estimate of the barrier height. To maintain the experimental structure, each bootstrap iteration consisted of resampling, with replacement, the observed data for each unique value of $\lambda$, thereby ensuring that the number of replicates per noise setting was preserved. For every resampled dataset, the polynomial regression model, using the previously determined optimal degree, was refitted, and the intercept coefficient was recorded. In Figure~\ref{fig:bootstrap_intercepts}, the empirical distributions for the `Fit first' and `Diff first' methods are presented.
The `Fit first' distribution is derived by generating bootstrap samples of the intercepts for the left and middle systems, respectively, and taking their difference, whereas the `Diff first' distribution is derived by generating bootstrap samples of $\Delta E$ directly.  The empirical distribution of the intercept values across all bootstrap replicates was subsequently used to quantify the uncertainty in this parameter. To capture the interval corresponding to $\pm 1 \sigma$, the $15^\mathrm{th}$ and $85^\mathrm{th}$ percentiles of this distribution were used, while the median value served as a robust point estimate for the intercept. These results are summarized in Table~\ref{tab:bootstrap_estimates} for both extrapolation methods. Excellent agreement is observed when comparing these bootstrap-based values with those obtained from estimation of the standard error of the intercept, as described in Sec.~\ref{sec:err-mitigation} and presented in Table~\ref{tab:summary-aqc}.
\begin{table}[h!]
    \centering
    \caption{Summary of ZNE extrapolation results using bootstrap estimation. All values are in mHa.}
    \begin{tabular}{lccc}
        \toprule
        Extrapolation method & Median & 15th percentile & 85th percentile \\
        \midrule
        Fit first & 24 & 15 & 32 \\
        Diff first& 17 & 14 & 21 \\
        \bottomrule
    \end{tabular}
    \label{tab:bootstrap_estimates}
\end{table}
\end{document}